\documentclass[aps,pra,showpacs,twocolumn,groupedaddress]{revtex4}
\newcommand \beq{\begin{eqnarray}}
\newcommand \eeq{\end{eqnarray}}
\usepackage{amsmath}
\usepackage{amssymb}
\usepackage{graphics}
\usepackage{bm}
\usepackage{graphicx}
\usepackage{subfigure}
\def\simge{\mathrel{%
       \rlap{\raise 0.511ex \hbox{$>$}}{\lower 0.511ex \hbox{$\sim$}}}}
\def\simle{\mathrel{
       \rlap{\raise 0.511ex \hbox{$<$}}{\lower 0.511ex \hbox{$\sim$}}}}
\usepackage{graphicx}

\date{\today}

\begin{document}
\title{Calculating energy shifts in terms of phase shifts }
\author{Zhenhua Yu,$^a$ Gordon Baym,$^{a,b}$ and C.\ J.\ Pethick,$^{a,c}$}
\affiliation{\mbox{$^a$The Niels Bohr International Academy, The
Niels Bohr Institute,}\\
\mbox{Blegdamsvej 17, DK-2100 Copenhagen \O,
 Denmark}\\
\mbox{$^b$Department of Physics, University of Illinois, 1110
  W. Green Street, Urbana, IL 61801} \\
\mbox{$^c$NORDITA, Roslagstullsbacken 23, SE-10691 Stockholm, Sweden} \\
}

\date{\today}

\begin{abstract}

 To clarify the relation of energy shifts to scattering phase shifts in
one-body and many-body problems, we examine their relation in a number of
different situations.  We derive, for a particle in a container of arbitrary
shape with a short-range scattering center, a general result for the energy
eigenvalues in terms of the s-wave scattering phase shift and the eigenstates
in the absence of the scatterer.  We show that, while the energy shifts for a spherical
container are proportional to the phase shift over large ranges, those for a cubic container have a more complicated
behavior.   We connect our result to the description of energy shifts in
terms of the scattering T-matrix.  The general relation is
extended to problems of particles in traps with smoothly varying potentials,
including, e.g., the interaction of a small neutral atom with a Rydberg atom.
We then consider the many-body problem for particles with a two-body
interaction and show that the energy change due to the interaction is proportional to an average of a
generalized phase shift that includes the effects of the medium.  Finally, we discuss why, even though individual energy levels are very sensitive to boundary conditions, the energy of a many-body system is not.

\end{abstract}
\maketitle

\section{Introduction}

    The art of calculating energies of many-particle systems in terms of
scattering phase shifts has a long history, dating back to Lenz \cite{lenz},
who derived the low-energy effective interaction between particles in terms of
the scattering length for two-body scattering, and to Beth and Uhlenbeck
\cite{beth}, who expressed the second virial coefficent for a classical gas in
terms of the phase shifts for two-body scattering.  Such a procedure is
convenient, since it separates the physics of the short-range scattering
process from that of single-particle motion in the external potential
confining the particles.  There are many variants of the formalism, including
the pseudopotential approach of Huang and Yang \cite{hy}.

    Further applications, particularly to systems in finite geometry, include
the early example of calculation of energy shifts of electronic
states due to impurities in a metal \cite{fumi}, and studies of
ultracold atomic gases confined in traps.  Yet further examples are
calculations of \mbox{Rydberg} molecules \cite{greene}, the
calculation of the energy levels of two interacting particles in a
harmonic trap \cite{harmonic1,harmonic2}, and the construction of
generalizations of the Gross-Pitaevskii equation to include
interactions beyond the usual pseudopotential proportional to the
s-wave scattering length between the particles \cite{gao,pethick}.
From a quite different perspective, L{\"u}scher \cite{luscher2}
derived a formula for the energies of bound states of two particles
in a box (with periodic boundary conditions) that is enjoying use in particle physics to extract scattering properties of
elementary particles from lattice calculations of energies
\cite{savage}.

Our primary aim in this paper is to clarify the relation between energy
shifts arising from interparticle interactions and the scattering
phase shifts.  We begin by considering the energy levels of a particle in a spherical box in the presence of a scatterer at the center of the box.  For simplicity, we focus on energies low enough that only s-wave scattering from the scatterer need be taken into account.  In the
presence of a short-range scatterer, the energy shifts are linear in
the s-wave phase shift, $\delta$, and not a trigonometric function
of $\delta$, such as the real part of the scattering amplitude.
However, in the case of a particle in a more generally shaped
container, with more complicated boundary conditions on the wave
functions, we find that the energy shifts depend
implicitly on $\tan \delta$ (see Eq.~(\ref{lu_0})).
In particular, as we shall see explicitly for a particle in
a cubic container, while the energy shifts are approximately
linear in $\delta$ for small $\delta$, the slope of the shift varies
from state to state. In the many body problem, as we illustrate with
a simplified separable interaction, a delta-shell potential, the
energy shift is linear in an effective scattering phase shift
that takes into account effects of the medium.

    We begin, in Sec.~II, by reviewing the energy shifts of a particle in a
spherical container.  We then go on in Sec.~III to derive a general expression
for the energy of a single particle moving in a container of arbitrary shape
in the presence of a scattering center.  The energy is obtained as an implicit
function of the s-wave phase shift of the scatterer and the eigenfunctions of
the one-particle problem in the absence of the scatterer.  We apply the
general result to the case of a cubic container in Sec.~IV.  In Sec.~V we
derive results for the cubic container from a T-matrix approach, and in
Sec.~VI describe applications to situations such as atoms in traps and
Rydberg atoms in which the particle moves in a potential that varies smoothly
in space, rather than in a container with infinitely repulsive walls.  Section
VII considers applications to the many-body problem, and discusses how the
formalism may be generalized to take into account the effects of the medium on
two-body scattering.  We take $\hbar=1$ throughout.

\section{Energy shifts in a sphere}

To set the stage, we first review the relation of the energy shifts
and s-wave scattering phase shifts for the familiar problem of a
particle of mass $M$ interacting with a short range
potential $v({\bf r})$  in a hard walled spherical container whose radius,
$R$, is much larger than the range, $b$, of the potential.
At low energies, scattering is predominantly s-wave, even if the potential is not central, and may be parametrized in terms of the s-wave phase shift, $\delta$.  Consequently, at low energies,   eigenstates of the energy are also eigenstates of the angular momentum, and we shall characterize states by the value of the angular momentum.  In this limit, states other than s-wave ones are unaffected by the scatterer.
For $v({\bf r})=0$, the single-particle  eigenenergies for s-wave states are
$E_\nu=k_\nu^2/2M$, with $k_\nu=\nu\pi /R$, and $\nu=1,2,3,\dots$,
with eigenfunctions
\begin{align}
 \psi_\nu(\mathbf r)=\frac1{\sqrt{2\pi R}}\frac{\sin(k_\nu r)}{r}.
\end{align}
In the presence of the scatterer, the eigenfunctions at points beyond the range of
the potential become
\begin{align}
 \psi'_\nu(\mathbf r)=A_1\frac{\sin(k'_\nu
   r+\delta(k))}{r},
 \label{wf}
\end{align}
where $A_1$ is a normalization constant.
The hard wall boundary condition implies that
\begin{align}
  k_\nu'R+\delta(k_\nu')=\nu\pi
\label{sd1},
\end{align}
or
\begin{align}
   \cot(k_\nu'R)=-\cot\delta(k_\nu').
     \label{sd2}
\end{align}
The shift in the wavevector, $\Delta k_\nu \equiv
k'_\nu-k_\nu=-\delta(k_\nu')/R$, is linear in the phase shift.
Without loss of generality we may take $\delta$ to lie in the range
$-\pi/2$ to $\pi/2$, since an increase of $\delta$ by $\pi$ takes
one from a given state in the sphere to the next higher state. 
Since beyond the range of the potential $ \psi'_\nu(\mathbf r)$ is a solution of the Schr\"odinger equation with zero potential, the
corresponding eigenenergies are $E_\nu=k_\nu'^2/2M$, and the shifts
in the energies due to the interaction are
\begin{align}
  \Delta E_\nu=\frac{{k'}_\nu^2-k_\nu^2}{2M}  = -\frac{2\pi\delta(k_\nu')}{Mk_\nu} \left(1-\frac{\delta(k_\nu')}{2k_\nu R}\right) |\psi_\nu(0)|^2.
  \label{compare}
\end{align}

The term quadratic in $\delta$ may be neglected if $\delta\ll1$, since $k_\nu R\gtrsim 1$.  Even for resonant scattering, $\delta \sim \pi/2$, the quadratic term may be neglected for states with many nodes ($k_\nu R\gg 1$). The energy shift is then given by
\begin{align}
  \Delta E_\nu  \approx -\frac{2\pi\delta(k_\nu')}{Mk_\nu}  |\psi(0)|^2.
  \label{compare}
\end{align}
Thus in these general situations, the energy shift is
linear in the phase shift (and not, e.g., the real part of the
scattering amplitude $\sim e^{i\delta}\sin\delta$).
The energy of a state must be determined self-consistently, since $\delta$ depends on energy, but if
$R^{-1}(|d\delta/dk|)\ll1$ one may replace $k'_\nu$ by the wave number of the state in the absence of the scatterer, $k_\nu$.

In the limit $k\to0$, away from a scattering resonance,
$\delta \to -ka_s$, with $a_s$ the s-wave scattering length.
Then Eq.~(\ref{compare}) gives
\begin{align}
  \Delta E(k)  = \frac{2\pi a_s}{M}  |\psi(0)|^2.
  \label{compare2}
\end{align}




\section{S-wave scattering in an arbitrary container}

    We turn now to establishing, for a particle of mass $M$
interacting with a short range central potential in a container of
arbitrary shape, the relation between the energy shifts of the
particle and the s-wave scattering phase shift. We consider a particle interacting with a scatterer located at some point within the container, which we take to be ${\bf r}= 0$.  We shall take the potential to be spherically symmetric and short-range, vanishing for $r>b$, and we
assume  that the
characteristic dimension of the container, $\lambda$,  is much
larger than the range of the potential, $b$.  We shall consider energies, $E$, sufficiently low that only
s-wave scattering is important. We denote the eigenfunction $\psi_k(\mathbf r)$ for energy
$E=k^2/2M$ by $\psi^<_k(\mathbf r)$ for $r<b$ and $\psi^>_k(\mathbf r)$ for $r>b$.
For distances $r\ll \lambda$,
$\psi^>_k(\mathbf r)$ may be expanded in terms of spherical waves as
\begin{align}
\psi_k^>(\mathbf r)=
A_2\Big[\frac{\sin(kr)}{kr}+&\tan\delta\frac{\cos(kr)}{kr}  \nonumber \\
+& \sum_{l=1}^\infty\sum_{m=-l}^l\alpha_{lm}j_l(kr)Y_{lm}(\theta,\phi) \Big] ,
\label{wf0}
\end{align}
where $j_l$ is the spherical Bessel function, the $\alpha_{lm}$ are expansion coefficients, and $A_2$ is a normalization constant.  In the sum, terms proportional to the spherical Neumann function $n_l$, which is singular for $r\to 0$, have been omitted, since we assume that only s-wave scattering is important, and consequently the wave function for other partial waves must be finite for $r\to 0$.
Thus asymptotically, in the region $b<r\ll 1/k$,
\begin{align}
\psi_k^>(\mathbf r)=A_2\left[\frac{\tan\delta}{kr}+1+{\cal O}(kr)\right].
\label{wfb_0}
\end{align}

To derive an expression for the energy of a state, we now find a second expression for the wave function, by expressing it in terms of the single
particle eigenstates  $\{\phi_n(\mathbf r)\}$ for $v({\bf r})=0$, with energy $E_n$ as follows.  The single
particle Green function for $v(\mathbf r)=0$ describing propagation of a particle from the origin to $\mathbf r$ is given by
\begin{align}
 G_k({\mathbf r},0)=\sum_{n}\frac{\phi_n(\mathbf r)
 \phi^*_n(0)} {E_n-E}.
  \label{g_0}
\end{align}
This satisfies the equation,
\begin{align}
  \frac1{2M}(\nabla^2+k^2)G_k(\mathbf r,0)=-\delta(\mathbf r),
  \label{ge_0}
\end{align}
which shows that $G_k(\mathbf r,0)$ satisfies the Schr\"odinger equation for $\psi_k^>(\mathbf r)$ when $\mathbf r\neq 0$.
 Thus outside the range of the
potential, $G_k(\mathbf r,0)$ satisfies the Schr\"odinger equation and in addition satisfies the same boundary conditions as the wave functions $\phi_n$ at the walls of the container for any boundary condition that is linear in the wave function.  Equations~(\ref{ge_0}) implies
that $G_k(\mathbf r,0)=M/2\pi r$ for $kr\ll 1$.
One can in fact show that for $r>b$
\begin{align}
   \psi_k^>(\mathbf r)=\frac{2\pi A_2\tan\delta}{Mk}G_k(\mathbf r,0),
  \label{nons_0}
\end{align}
by following Ref.~\cite{luscher2}. The function
\begin{align}
\chi_k(\mathbf r)=\psi_k^>(\mathbf r)-\frac{2\pi  A_2 \tan\delta}{Mk}G_k(\mathbf r,0)
\end{align}
is regular everywhere inside the container and is an eigenfunction with eigenenergy $E$ for $v(\mathbf r)=0$.
However, unless $E=E_n$, this is impossible, since the $\phi_n$ form a complete set, and therefore
$\chi_k(\mathbf r)=0$.
(For discussion of the more complicated case when  $E=E_n$, see Ref.~\cite{luscher2}.)

Matching the singular and nonsingular parts in Eqs.~(\ref{wfb_0}) and (\ref{nons_0}) for $kr\ll 1$, we obtain an expression for the energy of the state:
\begin{align}
 \frac{k}{4\pi}\cot\delta(k)=\lim_{r\to
  0}\left[\sum_{n}\frac1{2M}\frac{\phi_n(\mathbf r)
 \phi^*_n(0)} {E_n-E}-\frac1{4\pi r}\right],
\label{lu_0}
\end{align}
since non s-wave contributions vanish in the limit $r\to0$.

Equation~(\ref{lu_0}) is one of the main results of the paper, and determines the eigenenergies in the presence
of the potential. When the potential strength tends to zero, $E$ approaches an
unperturbed energy $\cal E$; accordingly Eqs.~(\ref{g_0}) and (\ref{nons_0}) implies that $\psi_k(\mathbf r)$ approaches the
state
\begin{align}
\phi^s_{\cal E}(\mathbf r) = C\sum_{E_n=\cal E}\phi_n(\mathbf r)\phi_n^*(0),
\label{ayw}
\end{align}
where the sum is over all possible degenerate states of energy
$\cal E$, and $C=\left(\sum_{E_n={\cal E}} |\phi_n(0)|^2\right)^{-1/2}$ is
the normalization constant. By construction, $\phi^s_{{\cal E}}(\mathbf
r)$ has a nonzero s-wave component, i.e., $\phi^s_{{\cal E}}(0)\neq 0$ if not all $\phi_n^*(0)$ equal zero.
The remaining states      $\phi^{\bar s}_{{\cal E}}(\mathbf r) $   in the subspace spanned by the states with $E_n={\cal E}$ vanish at ${\bf r}=0$, as we now show.
Since these states are energy eigenstates in the absence of the scatterer, we may write $\phi^{\bar s}_{{\cal E}}(\mathbf r) =\sum_{E_n={\cal E}}c_n\phi_n(\mathbf r)$.  We now evaluate the integral $\int d{\bf r} \left(\phi^s_{\cal E}(\mathbf r)\right)^*\phi^{\bar s}_{\cal E}(\mathbf r)$, which by orthogonality is zero if $s\neq \bar s$.    Inserting Eq.\ (\ref{ayw}) and the expansion for $\phi^{\bar s}_{{\cal E}}(\mathbf r)$ in the integral, one finds $\int d{\bf r} \left(\phi^s_{\cal E}(\mathbf r)\right)^*\phi^{\bar s}_{\cal E}(\mathbf r) =C\sum_{E_n={\cal E}} c_n \phi_n(0)=C\phi_{\cal E}^{\bar s}(0)   =0$.
 Therefore the $\phi^{\bar s}_{{\cal E}}(\mathbf r)$ for $s\neq \bar s$ vanish at the origin and these states will be unaffected when the scatterer at ${\bf r}=0$ is introduced.

For $k\to0$, away from a scattering resonance,
the left side of Eq.~(\ref{lu_0}) becomes $-1/4\pi a_s$, and so for $a_s$ small compared with a characteristic
dimension of the container,
$E$ must be very close to an unperturbed eigenenergy $E_n$.
Changing the order of taking the limit $r\to 0$ and summation in Eq.~(\ref{lu_0}),
we obtain
\begin{align}
\Delta E=E-E_n
=\frac{2\pi a_s}M |\phi^s_{E_0}(0)|^2;
\label{de}
\end{align}
thus the result (\ref{compare2}) found in a sphere holds for more
general geometry.

We now summarize the salient results of this section.  When the energies of states of a particle in the container in the absence of the scatterer are degenerate, one may construct one state, Eq.\ (\ref{ayw}), that does not vanish at $r=0$, while all other orthogonal states in the degenerate manifold vanish at $r=0$.  For weak scattering, $|\delta(k)|\ll \pi$, the energy shift of the first state is given by Eq.\ (\ref{de}) while the shift of the other states vanishes.  In our derivation of these results, we have taken into account degeneracy of the energy levels in the absence of the scatterer. For stronger scattering, it is necessary to solve Eq.\ (\ref{lu_0}).

The energy shift $\Delta E$ is reproduced by the expectation value of the effective interaction \cite{tony},
\begin{align}
   V_{\rm eff}(\mathbf r)=\frac{2\pi a_s}M\delta(\mathbf r),
   \label{effint_0}
\end{align}
in the {\em unperturbed} state, $\phi_n$.  In terms of the
interaction of two particles of mass $m$, the coefficient of the
effective interaction is $4\pi a_s/m$.

 One should note that the
effective interaction (\ref{effint_0}) is not useful for calculating
the perturbed wave function, $\psi_>(r)$, outside the range of the
potential in terms of a boundary condition at the origin.  Here one
needs rather to employ the pseudopotential \cite{hy}
\begin{align}
  V_{\rm ps}(\mathbf r)=-\frac{2\pi\tan\delta}{Mk}\delta(\mathbf r)
  \frac{\partial}{\partial r}r
\label{huang_0}
\end{align}
in the Schr\"odinger equation
\begin{align}
 \left(-\frac{\nabla^2}{2M} +V_{\rm ps}(\mathbf r)\right)\psi_>(r)=\frac{k^2}{2M}\psi_>(r).
  \label{ex_0}
\end{align}
The solution is clearly Eq.~(\ref{wf0}).  The present discussion is readily
generalized to higher angular momentum, with a pseudopotential involving
higher derivatives of the wave function \cite{feldmeier}.

\section{Energy shifts in a cubic box}

    We now apply Eq.~(\ref{lu_0}) to the specific example of scattering
by a spherically symmetric potential centered in a cubic box of side $L$, with
periodic boundary conditions.  This geometry, used in the calculations of
Refs.~\cite{luscher2,savage}, is more difficult to calculate analytically than
in a spherical container, since the boundary is incompatible with the
rotational symmetry of the central potential.  As one sees explicitly by
direct construction of the solutions of the Schr\"odinger equation, all
partial waves are mixed in determining the eigenenergies \cite{luscher2}.

   In the absence of the scatterer,  the states are simply
\begin{align}
  \phi_{\mathbf p}(\mathbf r)=\frac1{\sqrt{L^3}}e^{i\mathbf p\cdot\mathbf r},
\end{align}
where $\mathbf p=\{p_x,p_y,p_z\}=(2\pi/L)\{l_x, l_y,l_z\}$ with
$l_x,l_y,l_z=0,\pm \,1,\pm\,2, \dots$\;.  As before, we consider
only the s-wave contribution to the interaction, and from
Eq.~(\ref{lu_0}) find \cite{luscher2}
\begin{align}
 \frac{k}{4\pi}\cot\delta(k)=\lim_{r\to
  0}\left[\frac{1}{L^3}\sum_{\mathbf p} \frac{e^{i\mathbf p\cdot\mathbf
  r}}{p^2-k^2}-\frac1{4\pi r}\right].
\label{lu}
\end{align}

\begin{figure}
  \centering
  \subfigure{
    \label{cubek} 
    \includegraphics[width=3in]{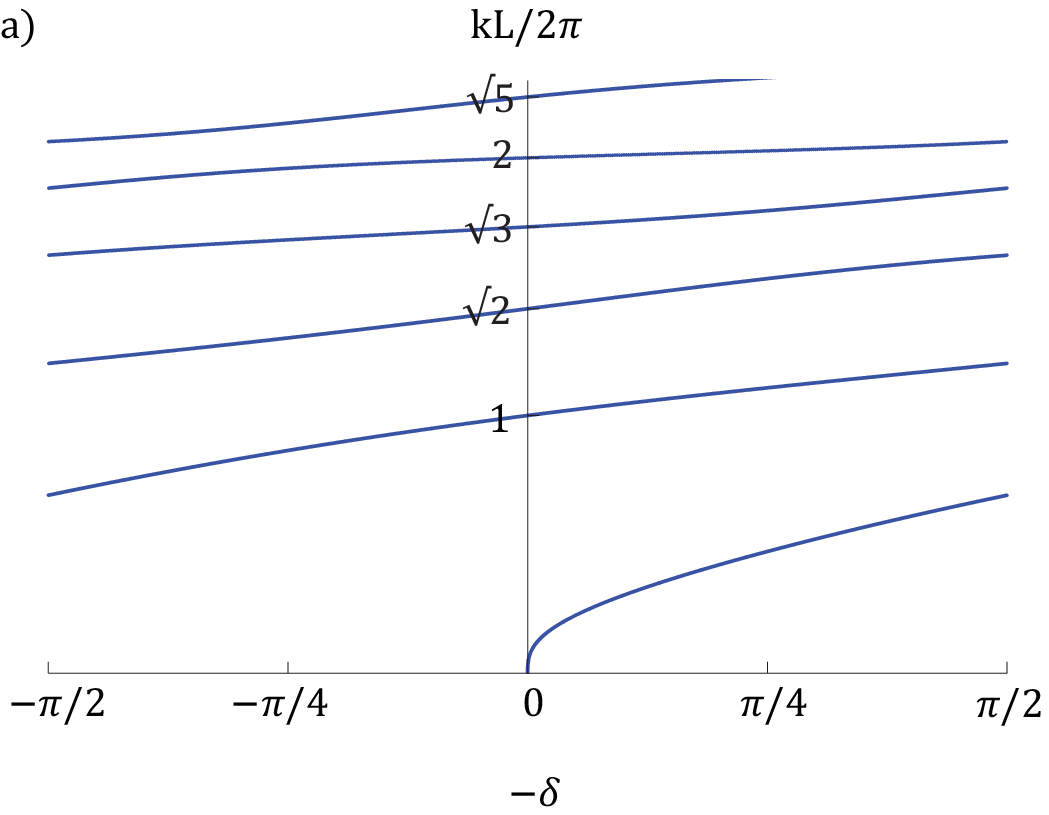}}
  \centering
  \subfigure
  {
    \label{cubee}
    \includegraphics[width=3in]{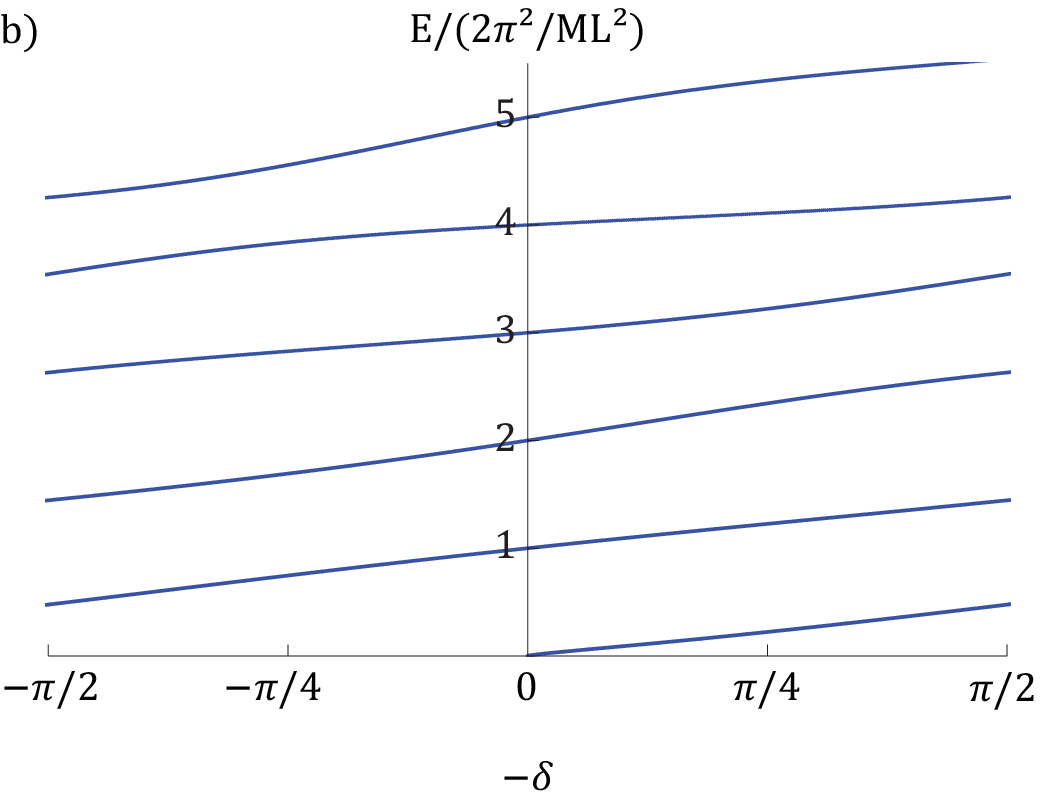}}
  \caption{(a) Wavenumbers $k$ and (b) eigenenergies $E$ vs. $-\delta$ in a cubic box.
  The curves are calculated from Eq.~(\ref{sa}) with $\Lambda=200\pi/L$.}

  \label{f1}
\end{figure}

    A relation similar to Eq.~(\ref{lu}) is more useful for
simulations on a lattice \cite{savage}:
\begin{align}
  \frac{k}{4\pi}\cot\delta(k)=\frac{1}{L^3}\sum_{|\mathbf p|<\Lambda}
  \frac{1}{p^2-k^2}-\frac{\Lambda}{4\pi},
\label{sa}
\end{align}
where $\Lambda$ is the momentum cutoff on the lattice.  To derive Eq.~(\ref{sa}),
we note that in the limit $r\to0$ the terms in the sum in Eq.~(\ref{lu}) for
$|\mathbf p|>\Lambda$ and $k \ll \Lambda$ can be written as
\begin{align}
  \frac{1}{L^3}\sum_{|{\mathbf p}| > \Lambda}
   \frac{e^{i\mathbf p\cdot\mathbf r}}   {\mathbf  p^2}
    = \int_0^\infty \frac{d^3p}{(2\pi)^3} \frac{e^{i{\mathbf p}\cdot {\mathbf
    r}}}{p^2} - \int_0^\Lambda \frac{d^3p}{(2\pi)^3}\frac{e^{i{\mathbf p}\cdot
   {\mathbf r}}}{p^2}  \nonumber\\
   =  \frac1{4\pi r} - \frac{\Lambda}{4\pi}.
\end{align}
For small $\delta$, the energy shift obeys Eq.~(\ref{de}).
Whereas the shift in $k$ for a sphere is linear in $-\delta$,
for a cube it is only approximately linear in
$-\delta$, as we see from Fig.~(\ref{f1}), which gives a plot of $k$ (panel (a)) and $E$ (panel (b)) vs.
$-\delta$ for the first few eigenstates, calculated from
Eq.~(\ref{sa}).   When $\delta$ is small (except in the ground
state),
\begin{align}
   k_l'L/2\pi \approx k_l L/2\pi -\alpha_{k_l}\delta,
\end{align}
where $E=k^2_l/2M$ is the eigenenergy in the
absence of the scatterer, and from Eq.~(\ref{sa}),
$\alpha_{k_l}=D_{k_l}/(k_lL)^2$, with $D_{k_l} = \sum_{\mathbf p}\delta_{p^2,k_l^2}$ the
degeneracy of the level $E$.
Therefore $\Delta k_l' L/2\pi\approx -\delta D_{k_l}/(k_lL)^2$
for small $\delta$, where $\Delta k_l'=k_l'-k_l$.  The linear relation is valid only
for small $\delta$, but with a slope which varies with the eigenstate of the cubic box.

 The different dependences of $k$ and $E$ on  $\delta$ for  cubic
and spherical boxes are due solely to the different boundary
conditions. The information on the boundary conditions is encoded in
Eq.~(\ref{lu_0}) through the unperturbed eigenvalues and eigenfunctions,
 while the effect of the short-range potential is
taken into account via $\delta$; in an arbitrary container, the energy change is a
function of $\cot\delta$ but the exact form of the function is
determined by the boundary conditions.

\section{T-matrix approach}

    With a view to generalizing results to situations other than dilute gases,
where binary collisions are the most important interaction process, it is
useful to derive the result (\ref{sa}) in terms of the T-matrix.  The
formalism may then be applied to calculate properties of interacting many-body
systems, when the elementary excitations are quite different from free
particles, e.g. liquid $^3$He at low temperatures.  Equation (\ref{lu_0}) was
derived by inspecting the Schr\"odinger equation with boundary conditions at
both small and large $r$.  The T-matrix formalism we give below is equivalent
to solving the Schr\"odinger equation in the one-body problem and its analogue
in the many-body problem to be discussed later provides a convenient way to
treat the interparticle interactions.

    The T-matrix is defined symbolically as
\begin{align}
   T(E)=v+v\frac{1}{E-H_0}T(E)=v+v\frac{1}{E-H}v,
  \label{t}
\end{align}
where $H_0$ is the free particle Hamiltonian, $v$ is the potential
operator, $H = H_0+v$, $(E-H)^{-1}$ is the particle propagator,
and $E$ the energy; this equation shows that the T-matrix
has poles in $E$ at the eigenenergies of the full Hamiltonian.
To derive Eq.~(\ref{sa}) we rewrite (\ref{t}) as:
\begin{align}
  T^{-1}(E)=v^{-1}- \frac{1}{E - H_0};
  \label{tinv}
\end{align}
with the on-shell T-matrix, at $E = k^2/2M +i\eta$, given by
\begin{align}
  T^{-1}(k^2/2M +i\eta) = -\frac{Mk}{2\pi}(\cot \delta(k) -i).
  \label{tdelta}
\end{align}

    For simplicity we assume a short ranged contact interaction,
$U\delta(\mathbf r)$ with a momentum cutoff, $\Lambda$.  Then in
free space Eq.~(\ref{tinv}) becomes,
\begin{align}
  T^{-1}(E)=U^{-1}- \int^\Lambda_0
   \frac{d^3p}{(2\pi)^3} \frac1{E-p^2/2M}.
\label{tcont}
\end{align}
Similarly in the cubic box,
\begin{align}
  T_{\rm box}^{-1}(E)=U^{-1} - \frac{1}{L^3}\sum_{|\mathbf p|<\Lambda}
  \frac{1}{E-p^2/2M}.
  \label{tbox}
\end{align}
Subtracting Eq.~(\ref{tbox}) from Eq.~(\ref{tcont}), evaluating both sides
at $E = k^2/2M +i\eta$, where $k^2/2M$ is now an eigenenergy in the box,
using Eq.~({\ref{tdelta}}) and the vanishing of $T^{-1}_{\rm box}$ at
$k^2/2M$, we find
\begin{align}
  \frac{k}{4\pi}\left(\cot \delta(k) -i\right)
   = \int^\Lambda_0 \frac{d^3p}{(2\pi)^3} \frac1{k^2 +i\eta -p^2}
\nonumber\\
   -\frac{1}{L^3}\sum_{|\mathbf p|<\Lambda} \frac{1}{k^2+i\eta-p^2}.
\label{tdiff}
\end{align}
The integral on the right in the limit  $k\ll \Lambda$ is $-(\Lambda
+ik)/4\pi$.  Dropping the $i\eta$ in the sum, we recover Eq.~(\ref{sa}).

    To see that the general formalism leads to $\Delta k_n R=-\delta$ for a
particle in a spherical box (cf.  Eq.~(\ref{sd1})), we apply the
T-matrix formalism.  For s-waves in a spherical box, the analog of
Eq.~(\ref{tbox}) is
\begin{align}
  T_{\rm sph}^{-1}(E)=U^{-1} - \frac{1}{2\pi R}\sum_{p<\pi\nu_{\rm
           max}/R}  \frac{p^2}{E-p^2/2M},
  \label{tsph}
\end{align}
where the $p$'s take on the values $\nu\pi/R$, with $\nu =
1,2,3,\ldots,\nu_{\rm max}$.  By the same argument as for the cubic box, we
derive the result
\begin{align}
   k\cot\delta = -\frac2\pi\Lambda + 2\nu_{\rm max}
   - \frac{2k^2}{R}\sum_{p<\pi\nu_{\rm max}/R} \frac1{k^2-p^2}.
\end{align}
Using
\begin{align}
  \sum_n\frac1{z-n\pi} = \cot z,
\end{align}
we obtain
\begin{align}
  \cot\delta(k)=-\cot(kR)+\frac1{kR}-\frac2{kR}\left(\frac{\Lambda
      R}{\pi}-\nu_{\rm max}\right).
  \label{trw}
\end{align}
Thus to recover the correct relation, $\Delta k R=-\delta$, we need to
choose the continuum cutoff $\Lambda$ to be $\pi(\nu_{max}+1/2)/R$, i.e., half
way between successive $\nu$ values.  This subtlety shares a similar origin as
in the transformation of a sum of $1/n$ into an integral leading to Euler's
constant $\gamma\approx0.577$.  An alternative way of dealing with this
problem is to introduce a smooth, rather than a sharp, cutoff.

    To further illustrate the T-matrix approach, we consider the delta
shell potential
\begin{align}
v(r)=(v_0/4\pi r_0^2)\delta(r-r_0),
\end{align}
which has the advantage that T-matrix can be constructed explicitly
without the need for a cutoff.  In the continuum limit, where we use as a
basis the s-wave functions,
\begin{align}
 \langle{\mathbf r}|k\rangle=\frac{\sin(kr)}{kr},
\end{align}
the T-matrix equation (\ref{t}) becomes,
\begin{align}
  T_{kk'}(E)=v_{kk'}+ \int \frac{d^3p}{(2\pi)^3}
       \frac{v_{kp}T_{pk'}(E)}{E-p^2/2M}.
\end{align}
The matrix elements of the delta shell potential are separable:
\begin{align}
  v_{kk'}=v_0\frac{\sin(kr_0)}{kr_0} \frac{\sin(k'r_0)}{k'r_0} \equiv v_0u_k u_{k'};
\end{align}
thus
\begin{align}
T_{kk'}^{-1}(E)= v_{kk'}^{-1}\left[
  1 -v_0\int_0^\infty\frac{d^3p}{(2\pi)^3}
     \frac{u_p^2}{E - p^2/2M}\right].
  \label{tb}
\end{align}
On shell ($k=k'$ and $E = k^2/2M + i\eta$),
\begin{align}
  T_{kk}(k^2/2M+i\eta) =\, \frac{v_0\sin^2(kr_0)}{k^2 r_0^2+
     iMkv_0\left(1-e^{2ikr_0}\right)/4\pi}.
\end{align}
Comparing with Eq.~(\ref{tdelta}) we find with some trigonometry that the
s-wave phase shift for the delta-shell potential is given by
\begin{align}
 \cot\delta(k)=-\frac{2\pi kr_0^2}{Mv_0}[1+\cot^2(kr_0)]-\cot(kr_0).
  \label{phase}
\end{align}

    Similarly in a sphere of radius $R$, with the s-wave functions
\begin{align}
  \langle r|n\rangle = \frac{1}{\sqrt{2\pi R}}\frac{\sin(k_nr)}{r}
\end{align}
as a basis, where $k_n=n\pi/R$, the T-matrix equation (\ref{t}) becomes,
\begin{align}
  T_{mn}(E)=v_{mn}+\sum_{l=1}^\infty
     v_{ml}\frac{1}{E-E_l}T_{ln}(E);
     \label{teqn}
\end{align}
the matrix elements of the delta-shell potential are now
\begin{align}
  v_{mn}=\frac{v_0}{2\pi Rr_0^2}\sin(k_m r_0)\sin(k_n r_0).
\end{align}
The solution of Eq.~(\ref{teqn}) is \cite{comparison}
\begin{align}
T_{mn}^{-1}(E) =  v_{mn}^{-1}\left[
       1-\frac{v_0}{2\pi Rr_0^2}\sum_{l=1}^\infty\frac{\sin^2(k_lr_0)}{E-k_l^2/2M}\right].
    \label{cpr}
\end{align}

    For $E=k^2/2M$,
\begin{align}
 \sum_{l=1}^\infty \frac{\sin^2(k_lr_0)}{k^2-k_l^2}
   =\frac{R}{2k}\sin^2(kr_0)[\cot(kR)-\cot(kr_0)],
\end{align}
so that
\begin{align}
 T_{mn}^{-1}  = v_{mn}^{-1}\left[
 1-\frac{Mv_0}{2\pi kr_0^2}\sin^2(kr_0)[\cot(kR)-\cot(kr_0)]\right].\label{co}
\end{align}
Comparing Eq.~(\ref{co}) with Eq.~(\ref{phase}), we find that at the pole of $T_{mn}$, at eigenenergy $E=k^2/2M$,
the expected result
\begin{align}
   \cot(kR)=-\cot\delta(k)
   \label{f};
\end{align}
this model calculation directly reproduces Eq.~(\ref{sd2}) (with $k=k_\nu^\prime$) within the T-matrix formalism for
the delta shell potential, avoiding the subtle issue of the cutoff $\Lambda$
in Eq.~(\ref{trw}).

\section{Particles in traps and Rydberg atoms}

So far we have considered containers in which the potential is zero
outside the range of the scatterer.  The results may be extended to
particles in a potential,  $V({\bf r})$, such as in an atomic trap.
If the scatterer is at point ${\bf r}_s$, the Green function for
the particle in the potential satisfies the equation
\begin{align}
  \left[-\frac1{2M}(\nabla^2+k^2) +V({\bf r})-V({\bf r}_s)\right]G_k(\mathbf r)=\delta({\mathbf r}_s),
  \label{ge_0trap}
\end{align}
where $k^2=2M[E-V({\bf r}_s)]$.  Provided $V({\bf r})-V({\bf r}_s)$ is
small compared with other energy scales in Eq.\ (\ref{ge_0trap}) at distances
from the scatterer of the order of the range of the potential, the arguments
of Sec.~III still hold, with the origin shifted to ${\bf r}_s$.  Explicitly,
the condition is that $|V({\bf r})-V({\bf r}_s)| \ll 1/Mb^2 $ for $|{\bf
r}-{\bf r}_s|\lesssim b$, in addition to the usual one $kb\lesssim 1$, where
$b$ is the range of the scattering potential (not $V(\bf r)$).
Equation (\ref{lu_0}) thus holds for this more general case, provided the
$\phi_n$ are taken to be the eigenfunctions including the effects of the
potential $V({\bf r})$.

The general result (\ref{lu_0}) can be applied to calculate the
effective potential between a Rydberg atom and a small neutral atom.
In the absence of the neutral atom, the eigenstates $\phi_n(\mathbf
r)$ for the outermost electron are hydrogenic wavefunctions for the
Coulomb potential $V_c(r)$ exerted by the ion, where we take the
nucleus to be at the origin.  We place the neutral atom at position $\mathbf{r}_s$,
of magnitude much larger than $a_0$ (the Bohr radius). The
eigenenergies $E$ of the outmost electron of the Rydberg atom are
functions of $\mathbf{r}_s$, and thus act as an effective potential
between the neutral and Rydberg atoms.   As a consequence of the electron's interaction
with the neutral atom, the wave function of the electron in the
Rydberg atom acquires a form near $\mathbf r=\mathbf r_s$ that corresponds to an s-wave phase shift $\delta(k)$. Thus from
Eq.~(\ref{lu_0}), the effective potential $E(r_s)$ is determined by
\begin{align}
 \frac{k}{4\pi}\cot\delta(k)=\lim_{\mathbf{r}\to
  \mathbf{r}_s}\left[\sum_{n}\frac1{2m_e}\frac{\phi_n(\mathbf r)
 \phi^*_n(\mathbf{r}_s)} {E_n-E(\mathbf{r}_s)}-\frac1{4\pi |\mathbf{r-r_s}|}\right],
\label{ra}
\end{align}
where $k=\sqrt{2m_e[E-V_c(r_s)]}$ is the scattering wavelength and $m_e$ is
the electron mass.  The sum over $n$ includes both the discrete and
continuum states.  Solving this equation is an alternative route to the results of Ref. \cite{greene}
for the interaction of a neutral atom with a Rydberg atom.  When applied to the problem of  two particles in a harmonic potential interacting via a short-range potential, Eq.\ (\ref{ra}) reproduces the result of Ref.~\cite{harmonic1}.

\section{Energy shifts in many-particle systems}

    We give now an example that allows one to identify the quantity in the
many-body problem that plays the role of the phase shift in two-body
scattering, and that shows how the effects of the medium enter the scattering
process.  This section may be regarded as an illustration of the basic
formalism for describing thermodynamic properties of a many-body system in
terms of fully renormalized single- and many-particle propagators given by De
Dominicis and Martin \cite{DeDominicisMartin} and reviewed in Ref.\
\cite{Bloch}.  We employ the delta shell model and calculate its effect in a
dilute binary fermionic gas within the ladder approximation \cite{nozieres,
PSbook}.  We assume that the system contains equal numbers of fermions in
internal states $|1\rangle$ and $|2\rangle$ of equal mass $m$, with chemical potential
$\mu$.

    Differentiating the grand potential per unit volume,  $\Omega$, with respect to the coupling
constant $v_0$ gives
\begin{align}
    \frac{\partial \Omega}{\partial v_0} = -\frac1{4\pi r_0^2}&\int d^3 \rho\, \delta(\rho-r_0)\langle
  \psi_1^\dagger({\bm \rho})\psi_2^\dagger(0) \psi_2(0)\psi_1({\bm \rho}) \rangle \nonumber  \\
    &=\frac{1}{\beta V}\sum_{\omega, \mathbf p} e^{\omega\eta}
    \frac{B(p,\omega)}{1-v_0 B(p,\omega)},
   \label{dfdv}
\end{align}
where $\eta$ is a positive infinitesimal and two-particle propagator is
\begin{align}
  B(p,\omega) =
 &-\frac{1}{\beta V}\sum_{z,\mathbf q}\frac{u_q^2}
 {(z-\epsilon_\mathbf{p/2+q})
 (\omega-z-\epsilon_\mathbf{p/2-q})}\nonumber\\
 =&\int\frac{d^3q}{(2\pi)^3} u_q^2
 \frac{1-2f(\epsilon_{\mathbf{p/2+q}})}{\omega+2\mu -p^2/4m-q^2/m},
\label{t0}
\end{align}
$\epsilon_{\mathbf q}=q^2/2m-\mu$, $\omega$ is a bosonic Matsubara frequency, and $z$ a fermionic one, $f$ is the Fermi distribution function,
$\beta=1/T$ with $T$ the temperature (we set the Boltzmann constant to unity),  and $V$ is the volume.  Here we approximate
the single particle propagators by those for free particles. In the low
density limit, at zero temperature,
$1-v_0 B(0,\omega)$
 is proportional to the inverse T-matrix for the separable potential and is
defined in Eq.~(\ref{tb}).

    Integration of Eq.~(\ref{dfdv}) with respect to $v_0$ gives the change in the grand potential due to the
interaction:
\begin{align}
\Delta \Omega=-\frac{1}{\beta}\sum_{\omega}\int \frac{d^3p}{(2\pi)^3}
e^{\omega \eta} \ln\left[t(p,\omega)/v_0\right], \label{t1}
\end{align}
where
\begin{align}
t(p,\omega) = \frac{v_0}{1-v_0B(p,\omega)}
\end{align}
is related to the T-matrix in the medium by $T_{kk'}(p,\omega)=u_ku_{k'}t(p,\omega)$,  with $u_k = \sin kr_0/kr_0$.

We convert the sum in Eq.~(\ref{t1}) into an integral, choosing the
branch cut of the logarithm to lie along the negative real axis. We
assume that $1+mv_0/4\pi r_0\ge0$ so that there is no two-body bound
state, and that the temperature exceeds the BCS transition temperature.  Then
the only singularity of $t(\mathbf p,\omega)$ is the branch cut
along the real $\omega$ axis, so that
\begin{align}
\Delta \Omega =-\int \frac{d^3p}{(2\pi)^3}\int_{-\infty}^\infty
\frac{d\omega}{\pi} f_B(\omega)\delta_m(p,\omega) \label{arg},
\end{align}
where $f_B$ is the Bose distribution function. The quantity \beq
\delta_m(p,\omega)=\arg(t(p,\omega+i\eta)) \eeq is the analog of
$\delta$ in the two body problem (cf. Eq.~(\ref{tdelta})).  Therefore
\begin{align}
\Delta \Omega =\int
\frac{d^3p_1}{(2\pi)^3}\frac{d^3p_2}{(2\pi)^3}    f(\epsilon_{{\bf
p}_1}) f(\epsilon_{{\bf p}_2}) v_{\rm eff}({\mathbf p}_1, {\mathbf
p}_2).  \label{phaseshiftE}
\end{align}
By the theorem of small increments, small changes in thermodynamic potentials keeping the natural variables appropriate to the chosen potential are the same \cite{LL}.  Consequently, Eq.\ (\ref{phaseshiftE}) gives the change in the energy density for fixed particle number and entropy per unit volume.
Equation (\ref{phaseshiftE}) has the usual form for the change in the energy density for a two-body interaction, with an
effective two-body interaction given by
\beq v_{\rm eff}({\bf p}_1,
{\bf p}_2)=\frac{\delta_m(|{\bf p}_1+{\bf p}_2|, \epsilon_{{\bf
p}_1}+\epsilon_{{\bf p}_2})}{{\rm Im} B( |{\bf p}_1+ {\bf p}_2|,
\epsilon_{{\bf p}_1}+\epsilon_{{\bf p}_2}+i\eta)}. 
\eeq 
In low density limit, $-{\rm Im} B( |{\bf p}_1+ {\bf p}_2|,
\epsilon_{{\bf p}_1}+\epsilon_{{\bf p}_2}+i\eta)/\pi$ reduces to
$m |{\bf p}_1-{\bf p}_2|/8\pi^2$,
 the
density of states for a pair of atoms with total momentum $|{\bf
p}_1+ {\bf p}_2|$.  For small $|{\bf p}_1-{\bf p}_2|$, the phase shift is given in terms of the s-wave scattering length by $\delta\to - |{\bf p}_1-{\bf p}_2|a_s/2$.  The factor of two arises because in the problem of scattering from a rigid scattering center considered earlier, we worked in terms of $k$, the momentum of the scattered particle, while in the two-body problem, the corresponding quantity is $ |{\bf p}_1-{\bf p}_2|/2$, the magnitude of the momentum of one of the particles relative to the center of mass of the pair.   The effective interaction in momentum space then reduces to the well-known result $4\pi a_s/m$ for two particles of equal mass (see Eq.\ (\ref{effint_0}) and the remarks following).

The many-body calculation above shows that the total energy change
is linear in a generalized phase shift that includes the effects of
the medium on the scattering process.
By contrast, Eq.~(\ref{lu_0}) shows that in an arbitrary
container, the energy changes of one-particle eigenstates in the presence of a short-range scatterer are
generally not linear in the s-wave phase shift, and one similarly expects  the energy eigenstates of two or more particles with binary interactions in a container of arbitrary shape not to be linear in the phase shift for the two-body interaction.  How does one account for this difference?
The essential
difference between the two situations may be seen by noting that the
energy of a (non-interacting) Fermi gas in the presence of a central
scatterer is a sum of contributions from a large number of
individual single-particle energy levels.  When the number of
particles $N$ is large compared with unity, the dominant
contribution to the total energy is insensitive to the behavior of
any particular energy level.  This is true even though the
single-particle energies are not linear functions of $\delta$, but
exhibit the undulating behavior shown in Fig.~\ref{f1}.
Contributions beyond the leading term are, however, sensitive to the
shape of the container and are responsible for, e.g., shell
structure in nuclear binding energies.  Similarly, in the many-body
problem with two-body scattering described above, the total energy
is insensitive to the boundary conditions a single-particle state satisfies on the walls of the container.

\section{Concluding remarks}

One of the main results of this paper is a generalization of an expression due to  L\"uscher \cite{luscher2} that expresses energy shifts of states in terms of scattering properties.  In particular, we have derived a general expression, Eq.\ (\ref{lu_0}) for the shift of the energy of the state of a single particle in a container of any shape external potential when a scatterer of short range is added.  In addition, a related result for a particle in a potential that is slowly varying in space was derived, Eq.\ (\ref{ra}).   The numerical results obtained for a particle in a cubic box show that the energies of low-lying states are a rather complicated function of the phase shift.  However, the functional form is constrained by the fact that when the phase shift is reduced from $\delta$ to $\delta - \pi$, the state of the system is transformed into the next higher state for a phase shift $\delta$.

In Sec.\ VII, we considered the many-body problem and showed that, for a model system of two species of fermion interacting via a two-body potential, the shift in the energy may be expressed as a {\it linear} function of an effective interaction proportional to a generalized phase shift that takes into account the effect of the medium on the scattering process.  The reason for this simple behavior for the many-body problem is that the free energy is a thermodynamic quantity, and this is sensitive mainly to the volume of the container, but insensitive to the details of the single particle wave functions, which depend on the shape of the container.

\section*{Acknowledgements}

    We are grateful to David Kaplan for calling our attention to the
applications of the L\"uscher formula in lattice simulations.  This research
was supported in part by NSF Grant PHY07-01611.


\begin{thebibliography}{20}
\bibitem{comparison} The delta shell potential is closely related to the
contact pseudopotential $U\delta(\mathbf r)$ with cutoff $\Lambda$, with the
identification $U\sim v_0$ and $\Lambda\sim 1/r_0$.  The factor $(\sin
k_lr_0)^2$ in Eq.~(\ref{cpr}) at small $k_r$ is $\sim k_l^2$, as in
Eq.~(\ref{tsph}); however the $k_l^2$ behavior flattens out at $k_l \sim
1/r_0$, regularizing the sum.
\end{thebibliography}
\end{document}